УДК 339.35

# ТРАНСФОРМАЦИЯ ЦЕПЕЙ ПОСТАВОК В СЕТЕВОЙ ЭКОНОМИКЕ: ПРИОРИТЕТЫ И ПЕРСПЕКТИВЫ
Калужский М.Л.

***Аннотация***: Развитие сетевой экономики меняет институциональное содержание экономических отношений. На смену традиционным каналам товародвижения приходят виртуальные сети распределения продукции. В статье раскрываются особенности и механизмы трансформации цепей поставок в электронной коммерции.
***Ключевые слова***: электронная коммерция, экзогенный подход, институциональная теория, интернет-маркетинг, распределительная логистика, сетевая экономика, управление маркетингом.

# TRANSFORMATION OF CHAINS OF DELIVERIES IN NETWORK ECONOMY: PRIORITIES AND PROSPECTS
Kaluzhsky M.L.

***Abstract***: Development of network economy changes the institutional maintenance of economic relations. On change to traditional channels of distribution virtual networks of distribution of production come. In article features and mechanisms of transformation of chains deliveries in e-commerce reveal.

Бурное развитие сетевой экономики в текущее десятилетие неизбежно сопряжено с глубокой трансформацией каналов распределения продукции. Интернет нивелирует расстояния, снижает трансакционные издержки и делает возможными мгновенные бизнес коммуникации в интерактивном режиме. В результате традиционные каналы товародвижения стремительно утрачивают свою конкурентоспособность. Сегодня мы стоим на пороге глубоких структурных изменений, благодаря которым трансформация цепей поставок уже в ближайшие годы изменит институциональную картину, как мировой, так и российской экономики.

Количественные проявления этого процесса пока отстают от качественных изменений, но скорость формирования сетевой инфраструктуры поставок и скорость распространения электронной коммерции позволяет говорить о нём как о новом источнике экономического развития [7]. Именно поэтому институциональный анализ трансформации поставок в электронной коммерции приобретает стратегическое значение при принятии управленческих решений, как на уровне корпоративного управления, так и на уровне государственного регулирования экономики.

**Предпосылки структурных изменений**. Структурные изменения в товародвижении, связанные с электронной коммерцией, возникают объективно в результате реализации возможностей, представляемых участникам рынка сетевой экономикой. Первичен здесь потребительский спрос, создающий благоприятную маркетинговую среду для внедрения инноваций. Можно предложить следующую классификацию предпосылок возникновения и развития структурных изменений под влиянием электронной коммерции:

1. *Внешние и внутренние предпосылки*. Внешние предпосылки структурных изменений заключаются в наличии потребительской готовности приобретать товары в условиях повышенного коммерческого риска по более низким ценам. Кроме того, немаловажную роль играют производители, обеспечивающие отгрузку товаров по виртуальным контрактам. Внутренние предпосылки структурных изменений заключаются в потенциальной возможности извлечения повышенной прибыли за счёт исключения из торговой цепи тра-диционных посредников (оптовое и розничное звенья), а также в наличии соответствующей инфраструктуры (платежи, логистика, торговые площадки).

2. *Макро и микроэкономические предпосылки*. Макроэкономические предпосылки структурных изменений заключаются в институциональных условиях, которые государство задаёт для экономической деятельности. Например, огромным стимулом для развития электронной коммерции в России стало увеличение беспошлинной стоимости ввозимых из-за рубежа товаров до 1000 евро в месяц на человека с вступлением в силу Таможенного кодекса Таможенного союза. Микроэкономические предпосылки структурных изменений заключаются в благоприятной экономической конъюнктуре, сложившейся на потребительском рынке. Неспособность традиционных участников рынка сформировать конкурентоспособную альтернативу стало мощным стимулом для возникновения и развития электронной коммерции.

3. *Экзогенные и эндогенные предпосылки электронной коммерции*. Экзогенные предпосылки структурных изменений определяются логикой развития сетевой экономики и интернет-технологий. Появление в электронной коммерции новых технологических возможностей (высокоскоростной Интернет, торговые и платёжные сервисы и т.п.) неизбежно ведёт к появлению новых форм и методов товародвижения. Эндогенные предпосылки структурных изменений определяются логикой развития малого предпринимательства в Интернете, не обладающего ресурсами для ведения традиционного бизнеса.

4. *Естественные и искусственные предпосылки*. Естественные предпосылки структурных изменений формируются условиями спроса и предложения на потребительском рынке. Два основных фактора, выступающие в качестве детерминанты развития электронной коммерции – резкое снижение трансакционных издержек и падающая покупательная способность населения. Искусственные предпосылки структурных изменений формируются в результате институциональной политики государства и решений, принимаемых крупными участниками рынка. Например, государственная поддержка почтовой логистики при организации розничных интернет-продаж в Китае стимулировала рост электронной коммерции и экспансию китайских товаров по всему миру.

5. *Объективные и субъективные предпосылки*. Объективные предпосылки структурных изменений представляют собой объективные параметры рыночной среды, определяющие экономическую эффективность применения электронной коммерции. К ним можно отнести, например, количество интернет-пользователей, степень насыщения рынка, уровень развития торговой, платёжной и транспортной инфраструктуры электронной коммерции. Субъективные предпосылки структурных изменений являются следствием управленческих решений, принимаемых на двух основных уровнях институционального управления – государственном и корпоративном. Речь здесь может идти о влиянии субъективных факторов в виде нормативного регулирования электронной коммерции или о стратегических приоритетах крупных провайдеров логистических услуг.

6. *Общие и частные предпосылки*. Общие предпосылки структурных изменений в товародвижении представляют собой совокупность факторов, определяющих условия функционирования институтов сетевой экономики и электронной коммерции в целом. Современный электронная коммерция возможен только в условиях сетевой экономики, вне которой он утрачивает свою конкурентоспособность и привлекательность. Частные предпосылки представляют собой совокупность факторов, относящихся к решению организационных проблем электронной торговли.

Виртуальность электронной коммерции служит причиной высокой степени её зависимости от уровня развития и качества работы соответствующей логистической инфраструктуры. Основное институциональное преимущество электронной коммерции заключается в существенной экономии на трансакционных издержках в сравнении с традиционной экономикой. В обычной коммерции трансакционные издержки торгового предприятия включают в себя затраты на закупку и хранение товара, содержание торгового персонала, ведение переговоров, заключение сделок, поставку товара и т.д. В электронной ком-мерции этих издержек просто не существует, так как процесс товародвижения в сети распределения виртуализируется и автоматизируется.

Возможности электронных торговых площадок, например, позволяют продавцам в автоматическом режиме формировать виртуальный ассортимент из тысяч наименований товара. Ведение переговоров и заключение сделок успешно заменяет функционал провайдеров услуг электронной торговли. Отгружает товар покупателям непосредственно сам поставщик. Поэтому традиционные оптово-розничные посредники выпадают из цепи товародвижения вместе с сопутствующими их деятельности трансакционными издержками, меняя технологическую структуру целого сектора экономики.

**Структурные изменения в цепях поставок**. В традиционной теории логистики под понятием «цепь поставок» (англ. *Supply chain*) понимается ряд «*видов деятельности и организаций, через которые материалы проходят во время своего перемещения от поставщиков начального уровня до конечных потребителей*» [10].

Подразумевается, что некто на свой страх и риск добывает сырьё, затем поставляет его производителю. Производитель на свой страх и риск производит конечный продукт и поставляет его торговым посредникам. Посредники на свой страх и риск пытаются продать продукт конечным потребителям.

Потребитель явно выпадает из этой схемы. Его мнение никого не интересует. Он не является субъектом экономических отношений. Он – всего лишь объект приложения маркетинговых усилий продавцов. Не случайно родоначальник современной теории менеджмента П. Друкер (P. Drucker) писал: «*цель бизнеса – создать потребителя, т.е. привлечь независимого внешнего субъекта, способного выбирать и готового заплатить за товар*» [2].

В такой модели маркетинговых взаимоотношений обратная связь потребителя с участниками цепи поставок чрезвычайно затруднена. Поэтому для стимулирования потребительского спроса продавцы изощряются в использовании маркетинговых инструментов в виде рекламных акций, скидок, пропаганды и т.п. Если товар удаётся продать, то вложения рано или поздно окупаются. Если товар не пользуется спросом – участники несут потери на всех этапах цепи поставок.

Аналогично в комплексе маркетинга (концепция «4P») есть всё для организации продаж (цена, товар, сбыт, коммуникации), но отсутствует конечный адресат – потребитель [3]. Вместе с тем, любой маркетолог знает, что массовые продажи способен обеспечить только массовый спрос. Если товар действительно востребован на рынке, то он будет продавать себя сам независимо от маркетинговых ухищрений продавца. Если же товар не востребован на рынке, то продавцу придётся приложить титанические усилия для его продвижения, эффект от которых будет крайне недолог.

Поэтому для электронной коммерции традиционная концепция организации товародвижения не подходит, так как классические эндогенные подходы в сетевой экономике здесь работают с точностью до наоборот, так как объём и параметры поставляемых товаров определяют потребители, а не производители и поставщики. Даже описать этот процесс с помощью эндогенных подходов невозможно, поскольку роль рыночных субъектов менее значима, чем роль потребительского спроса.

Экзогенную модель организации цепочки поставок можно обнаружить в набирающей популярность на Западе концепции управления цепью поставок (Supply Chain Management – SCM). Основная цель SCM – «*создание ценности для организаций, являющихся участниками цепи поставок, с особым акцентом на конечном потребителе*» [11]. При этом SCM традиционно принято относить исключительно к сфере B2B [11], что не совсем верно, так как и в сфере B2C она развивается вполне успешно.

Процесс организации цепочки поставок в SCM носит ярко выраженный экзогенный характер. Принцип действия цепочки поставок достаточно прост. Сначала потребители активно реагируют на предлагаемую рынку модификацию товара. Затем торговые посредники массово переключаются на продукцию определённого производителя. Конкурирую-щие производители начинают копировать популярные модели товара и рынок «разворачивается». И только тогда волна заказов на сырьё докатывается до поставщиков.

В результате потребитель из объекта маркетинга превращаются в его главного субъекта, отправную точку цепочки поставок. *«Единственным субъектом, который вкладывает деньги в цепь поставок,* – отмечают Р.Б. Хэндфилд и Э.Л. Николс, – *является конечный потребитель, который принимает решение о покупке продукта; все остальные лишь перераспределяют его деньги между участниками цепи поставок»* [11].

Для электронной коммерции это утверждение особенно актуально, так как здесь потребители практически напрямую взаимодействуют с производителями. Не случайно Ф. Котлер на вопрос о том, кто больше выиграет от развития Интернета – компании или потребители – даёт совершенно однозначный ответ: *«Больше всех выиграют потребители»* [3]. Эти потребители уже не идут на поводу у решивших всё за них производителей и торговых посредников. Наоборот, они в полной мере пользуются свободой выбора и самостоятельно принимают решения о покупке [12].

Если в традиционной экономике товарный ассортимент формирует торговый посредник и его поставщики, то в электронной коммерции царит совершенная конкуренция и товарный ассортимент изначально близок к абсолютному значению. Интернет делает одинаково доступным всё товарное предложение на рынке, независимо от страны происхождения и места расположения товара.

**Реорганизация цепей поставок**. Основная задача реорганизации цепей поставок в электронной коммерции заключается в том, чтобы находить *«новые способы автоматизации и ускорения своих бизнес-процессов»*, переходя от *«традиционных методов ведения бизнеса к электронной торговле, мгновенно реагируя на изменение потребностей клиентов и осуществляя необходимые изменения в своей деятельности при появлении новых возможностей на рынке»* [11]. Виртуальная среда Интернета в силу своей доступности представляет участникам рынка широчайшие возможности для реорганизации цепей поставок.

Вместе с тем, при кажущейся простоте, существуют достаточно серьёзные сложности с реорганизацией цепей поставок в электронной коммерции. Для производителей эти сложности связаны с решением двух видов логистических задач:

1) оптимизация производственного процесса;
2) организация информационных потоков.

Самостоятельно решить эти задачи не в состоянии ни один, даже самый крупный, производитель товаров. Однако именно от их решения зависит возможность эффективного организации цепей поставок.

1. ***Оптимизация производственного процесса*** подразумевает работу производителя «с колёс», когда объёмы производства определяются размером получаемых заказов, а складские запасы минимальны. Это сложнее, чем ставшая уже классической японская система «Канбан», так как здесь недостаточно организовать ритмичное производство без складских запасов. Тут требуется повышенная степень готовности производства к выполнению поступающих заказов. Проблема состоит в том, что даже среднесрочное производственное планирование в таких условиях теряет свою актуальность. Сокращение заказов вызывает немедленное обновление ассортимента, тогда как максимальный объём производства определяется производственными возможностями предприятия.

Как промежуточная форма организации поставок в электронной коммерции может применяться т.н. «предпродажа» (англ. *Pre-Sale*). В этом случае покупателю товар отгружается не сразу с момента оплаты заказа, а по завершении производственного цикла (обычно 10-15 дней). При большом количестве модификаций товара (например, по цвету) задача может решаться через массовое изготовление полуфабрикатов, а затем быструю доводку их до требований заказчика.

В такой модели организации производства наиболее конкурентоспособными будут являться производители с коротким производственным циклом. Причём их конкуренто-способность зависит не только от организации производства, но и от организации поставок сырья. Указанная форма отношений немыслима без развитой системы логистических посредников, принимающих на себя риски и ответственность за проведение рыночных трансакций.

2. ***Организация информационных потоков*** подразумевает не только инструменты электронных продаж и обмен информацией между участниками цепей поставок. Для производителей не менее важно обеспечить информационное сопровождение внутренних процессов, связанных с организацией производственного цикла (поставки сырья, производство и отгрузку продукции).

Помимо информации о состоянии рыночного спроса, фактором обеспечения конкурентоспособности здесь становится скорость принятия и исполнения производственных решений на основании полученной информации. «*Ускорение обмена информацией и её прямая связь с реальными событиями на рынке*, – отмечают М. Кристофер и Х. Пэк, – *в большей степени, чем что-либо другое, могут повысить конкурентоспособность компании*» [6].

Ещё недавно для отечественных предприятий решение этой задачи было неразрешимой проблемой, требующей отвлечения значительных ресурсов на компьютеризацию производства без гарантии положительного результата. Сегодня проблема решается с помощью провайдеров облачных сервисов, делающих ненужными покупку дорогостоящего оборудования и привлечение высококвалифицированных программистов.

В электронной коммерции в наибольшей степени происходит отделение информационных от материальных потоков, которое многие авторы определяют как один из важнейших сдвигов в мировой экономике [14]. Результатом такого сдвига становится смена приоритетов в управлении поставками – от оптимизации товародвижения к оптимизации информационных потоков. «*Ключевой целью управления цепочкой поставок,* отмечают М. Кристофер и Х. Пэк, – *является сбор информации о требованиях покупателей и потребителей, максимально приближенный к местам продаж или использования*» [6].

В результате для поставщиков продукции на смену складской, транспортной и торговой логистики приходит распределительная логистика. Всё остальное передаётся в доверительное управление специализированным провайдерам логистических услуг. «*Распределение полномочий внутри цепи поставок в условиях современного рынка основано на распределении информации*», – отмечают Р. Б. Хэндфилд и Э.Л. Николс [11]. Уже одно это свидетельствует о глубокой трансформации самой сущности цепей поставок.

**Реорганизация производства**. Переход к организации производственного процесса на основе широкого применения для сбыта продукции электронной коммерции требует кардинального изменения всей системы управления предприятием и перераспределения полномочий. Для этого необходимо пересмотреть основополагающие принципы организации экономической деятельности.

Известный американский институционалист Р. Коуз в своей книге «Фирма, рынок и право» пишет: «… *фирмы должны возникать просто для осуществления действий, которые в противном случае совершались бы через рыночные трансакции (разумеется, если внутрифирменные издержки меньше, чем издержки рыночных трансакций)*» [5]. Особенность сетевой экономики и виртуальных трансакций заключается в том, что здесь издержки рыночных трансакций значительно меньше, чем внутрифирменные издержки.

В результате пропадает экономическое основание для существования крупных производственных корпораций, производящих товары массового спроса и выполняющих весь спектр производственных функций: от закупки сырья до организации товародвижения. На виртуальном рынке как грибы после дождя возникают сегодня сетевые компании, стремительно завоёвывающие рынок и вытесняющие с него традиционных производителей.

Противостоять традиционные товаропроизводители этому процессу не способны, так как конкуренция идёт в виртуальной среде, где преимущества на стороне виртуальных организаций. Единственный способ не проиграть в конкурентной борьбе для них заклю-чается в том, чтобы самим трансформироваться и стать частью инфраструктуры сетевой экономики [1]. Там, где речь идёт о производстве материальных, а не виртуальных товаров (программного обеспечения, информационных продуктов и т.д.) полная виртуализация бизнеса невозможна. Кто-то должен производить эти товары, поставлять сырьё и обеспечивать товародвижение.

Основная идея подхода, диктуемого сетевой экономикой, заключается в отказе от тотальной интеграции управленческих функций внутри одного экономического субъекта. Все производственные, распределительные и иные полномочия, которые могут более эффективно реализовываться внешними провайдерами, должны быть переданы внешним провайдерам. Время крупных предприятий, самостоятельно разрабатывающих, производящих и продвигающих на рынке товары массового спроса прошло. На смену ему приходит время разделения полномочий между независимыми экономическими субъектами на контрактной основе. Это новая реальность сетевой экономики, ведущее место в которой занимают электронная коммерция и виртуальные трансакции.

В качестве теоретического обоснования грядущих институциональных изменений в экономической практике можно использовать т.н. «концепцию расширенного предприятия», основанную на приоритете *«внешних источников конкурентного преимущества»* [6]. Эта концепция включает в себя три базовых принципа, определяющих институциональные особенности новых форм экономических отношений в сетевой экономике:

1) стратегический переход «*от функций к процессам*» – через приоритетное развитие межфункциональных связей как внутри компании, так и с её партнёрами, а также с потребителями.

2) стратегический переход «*от товаров к покупателям*» – через признание удовлетворённости покупателей в качестве товара основной целью коммерческой деятельности, выраженной в получаемой прибыли;

3) стратегический переход «*от прибыли к эффективности*» – через отказ от стратегии контроля прибыльности в пользу стратегии контроля трансакционных издержек.

Суть «расширенного предприятия» заключается в том, что в процессе организации поставок происходит взаимное делегирование не прав (на поставку сырья, продукции и т.д.), а взаимное делегирование ответственности. *«Использование информационных технологий для координации функций и целостного управления интегрированной деятельностью позволяет распределить ответственность за результаты работы в масштабах всей организации»*, – отмечают американские маркетологи Д. Дж. Бауэрсокс и Д. Дж. Клосс [1].

«Организация» при этом виртуализируется и перестаёт существовать как отдельный правовой субъект. Формальный товаропроизводитель больше не изготавливает самостоятельно продукцию и не продвигает её на рынке. Он является держателем технологий, организует и управляет процессом производства и продвижения товаров, трансформируясь в единый интеграционный центр. Остальные функции выполняют с меньшими издержками и большей эффективностью внешние контрагенты и провайдеры.

Д. Дж. Бауэрсокс и Д. Дж. Клосс выделяют пять стадий эволюционного развития логистического управления производственным процессом (см. Рис. 1) [1].

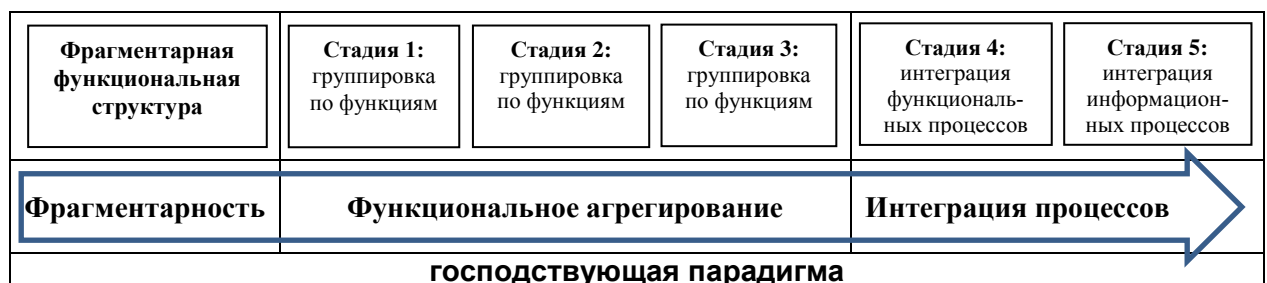

Рис. 1. Эволюционное развитие логистического управления

Первые три стадии непосредственно связаны с выполнением логистических операций и к сетевой экономике отношения не имеют. Четвёртая стадия связана с качественным переходом от управления функциями к управлению процессами. На этой стадии возрастает роль достоверной информации и происходит делегирование полномочий самостоятельным подразделениям и партнёрам компаний.

Непосредственно к сетевой экономике относится пятая стадия, связанной с трансформацией административно-командной иерархии в виртуальную организацию. Причём, термин «виртуальный» предлагается понимать как «*наличествующий по существу и по проявлениям, хотя и не получивший формального признания или статуса*» [1]. На этой стадии происходит разделение логистических функций между участниками «*электронной керитсу*», когда участники сосредотачиваются на выполнении отдельных логистических процессов.[1]

В экономической практике этот процесс идёт довольно давно. Его проявления можно увидеть, например, на упаковках товара в виде надписей: «*Производитель – компания «Х» (Япония). Страна-изготовитель – Китай*». В экономической теории также пришла пора разделить понятия «*производитель товара*» и «*изготовитель товара*». Развитие промышленных технологий неминуемо ведёт к универсализации и повышению гибкости производств, когда независимые изготовители массово производят товары по заказам их формальных производителей.

Основная задача здесь связана с решением проблемы максимального ускорения производственного процесса. Идеальной для эффективного применения электронной коммерции является ситуация, когда после получения оплаченных заказов производитель в течение считанных часов организует изготовление и отгрузку товара покупателю.

Пионерами в создании технологий такого рода производств стали японские товаропроизводители. Их преимущество заключалось в том, что «*японские компании концентрируются на гибкости и скорости, они способны выпускать товары уже на стадии прототипа, а затем быстро модифицировать их в зависимости от реакции потребителя*» [6]. Залогом этого служат сравнительно небольшие расстояния в Японии и развитая логистическая инфраструктура. Однако они не смогли изменить институциональную картину мировой экономики, так как не обладали преимуществами китайской дешевой силы.

Во Всемирной сети сегодня можно обнаружить массу аналогов такой организации производства. Глобальная сетевая инфраструктура формируется сегодня стремительными темпами и в виртуальных рамках сетевой экономики китайские компании изготавливают товары по спецификациям формальных производителей со всего мира. С другой стороны, производители больше не должны тратить ресурсы на поддержание загрузки производственных мощностей, покончив с необходимостью содержания персонала и излишних производственных мощностей.

В итоге изготовление конечной продукции постепенно трансформируется в логистический сервис, осуществляемый независимыми провайдерами. Одновременно формируется независимая инфраструктура, обеспечивающая реализацию заключаемых контрактов. Например, китайская торговая площадка Alibaba уже предлагает зарубежным производителям инспекционный сервис по проверке контрагентов, изготовляющих и поставляющих товары в соответствии с заключенными контрактами.[2] В качестве стандартного набора услуг предлагаются: первичный осмотр производства, выборочный контроль поставляемой продукции, контроль загрузки контейнера и производственный аудит. Аналогичные провайдерские услуги существуют и в России.[3]

**Реорганизация товародвижения**. Реорганизация товарных поставок при переходе к электронной коммерции не только позволяет значительно снизить трансакционные из-держки поставщика, но и требует коренной перестройки всего механизма товародвижения. Это связано с переходом к принципиально новой модели организации поставок, основанной на делегировании полномочий и ответственности независимым посредникам, а также на создании с их участием неформальных сетей товародвижения.

---

[1] *Керитсу (япон.)* – добровольное объединение независимых компаний на основе соглашения о разделении ответственности в общем проекте.
[2] Инспекционный сервис / Торговая площадка Alibaba. – http://inspection.alibaba.com/
[3] Сайт компании «ИСТЭКС». – http://eastex.ru/

Традиционная система товародвижения была немыслима без аккумуляции у торговых посредников большого количества товарных запасов и широкого использования банковского кредитования. Поэтому обращение субъектов электронной коммерции к традиционным каналам сбыта нивелировало их конкурентные преимущества при организации поставок. Традиционные продавцы получают определённые преимущества в процессе непосредственных продаж, но при этом вынуждены нести все издержки, связанные с содержанием каналов товародвижения.

Сетевая экономика, напротив, располагает своими уникальными каналами товародвижения, в полной мере раскрывающими трансакционные преимущества виртуализации сбытовых процессов. Эти каналы позволяют отказаться от накопления товарных запасов у торговых посредников и обойтись без привлечения банковского кредитования. Как совершенно справедливо отмечают Дж. Р. Сток и Д. М. Ламберт: «*наилучшие каналы создаются тогда, когда никакая другая группа институтов не обеспечивает более высокой прибыли или большей степени удовлетворения потребителя*» [8].

Традиционная система организации поставок обладает существенными недостатками, исправить которые возможно при помощи электронной коммерции:

1. Оптово-розничное звено объективно не в состоянии своевременно поставлять производителям информацию о потребительском спросе. В ассортименте розничного торговца сотни, а то и тысячи наименований товаров от множества производителям. Им просто некогда собирать маркетинговую информацию для каждого поставщика. В результате товаропроизводители месяцами не знают о судьбе поставленного товара и не могут планировать производство в соответствии с потребительским спросом.

2. Значительные финансовые ресурсы товаропроизводителей отвлекаются на товарное кредитование посредников и оплату услуг кредитных учреждений. В условиях отсутствия информации о реальных потребностях рынка потери от несоответствия структуры спроса и структуры производства неизбежны. Производители вынуждены перекладывать эти издержки на потребителей через увеличение отпускных цен, что значительно снижает конкурентоспособность продукции.

3. Независимое оптово-розничное звено не заинтересовано в продаже товаров конкретных товаропроизводителей, ориентируясь в первую очередь на свои коммерческие интересы. В условиях рыночной конкуренции товаропроизводители часто становятся жертвами оппортунистического поведения торговых посредников, перекладывающих на них убытки от собственных коммерческих просчётов и промахов.

4. Отсутствие у товаропроизводителей актуальной информации о рыночном спросе и его конъюнктуре ведёт к росту издержек на продвижение продукции. В условиях жесткой конкуренции высокорискованные затраты на продвижение брендов лишь увеличивают потери от неэффективных маркетинговых решений, перекладываемые на плечи потребителей. Это не позволяет традиционной системе товародвижения конкурировать с электронной коммерцией ни по ценам, ни по скорости адаптации к требованиям рынка.

В электронной коммерции в процессе товародвижения участвует только поставщик, торговый посредник и покупатель, либо поставщик и покупатель. Длинные каналы товародвижения здесь экономически не обоснованы, так как Интернет снимает ограничения, связанные с удалённостью рынков сбыта. Функции, связанные с логистическим обеспечением продаж, передаются независимым посредникам (провайдерам услуг), которые оказывают стандартные логистические услуги со стандартными параметрами качества.

Это значит, что поставщику товара не нужны промежуточные склады для хранения товарных запасов и удалённые представительства для расширения географии поставок. Ему не нужно обращаться к оптово-розничным посредникам, так как покупатель из друго-го региона ничем не отличается от покупателя из своего региона. Его основная функция заключается в том, чтобы получить электронный предоплаченный заказ и отгрузить товар через логистического посредника по стандартному контракту.

В маркетинговой логистике существует т.н. «правило квадратного корня», согласно которому «*сокращение товарно-материальных запасов пропорционально квадратному корню из числа мест расположения складов до и после рационализации*» [6]. Независимо от того, сколько было складов в традиционной системе распределения, электронная коммерция подразумевает наличие только одного склада – склада производителя (поставщика) товара. Склады транспортных провайдеров можно не учитывать, так они самостоятельно несут всю полноту ответственности за организацию поставок.

Статистика свидетельствует о том, что пока даже в США средние показатели оборачиваемости средств у традиционных поставщиков товаров массового спроса превышают полгода и более [6]. При этом многие компании из-за длинных неуправляемых торговых цепочек не в состоянии хоть как-то влиять на этот показатель. В электронной коммерции, напротив, оборачиваемость определяется характером текущего рыночного спроса, так как товар поставляется покупателям на условиях полной предоплаты. Подразумевается, что товары изготавливаются на высокотехнологичных производствах небольшими партиями и сразу же отгружаются заказчикам.

Использование электронных продаж в качестве отправной точки организации товародвижения ведёт не только к сокращению товарных запасов и уходу от диктата оптово-розничного звена. Электронная коммерция позволяет производителям контролировать поставки на всех этапах товародвижения, оптимизировать производство и получать информацию о потребительском спросе из первых рук. В традиционной логистике такое было бы невозможно. Основная проблема там заключается в решении задач по оптимальному распределению товарно-материальных ресурсов. Не случайно процессы интеграции цепей поставок в традиционной логистике охватывают лишь два вида логистических потоков: поток запасов и поток информации [1].

Не случайно на первое место в электронной коммерции выходят не потоки запасов и информации о товарных предложениях, а потоки платежей и информации о потребительском спросе. Институциональной базой здесь служит логистическая инфраструктура. Это – принципиальное отличие, обеспечивающее безусловное преимущество электронной коммерции перед традиционной торговлей за счёт ускорения оборота, экономии ресурсов и резкого повышения эффективности поставок товаропроизводителей.

Многие авторы отмечают сегодня, что в системе электронной коммерции «*традиционные службы снабжения (а в некоторых случаях и сбыта) становятся ненужными, их функции выполняет глобальная информационная система управления ресурсами предприятия*» [13]. При этом, готовые технологические решения, связанные с трансформацией производственных процессов на основе широкого применения облачных технологий, достаточно широко представлены как за рубежом, так и в России.

Однако их слабым местом является недостаточная проработанность вопросов маркетингового обеспечения виртуализации продаж в сфере B2C, хотя именно рынок продаж B2C доминирует сегодня в сетевой экономике. Уже недостаточно просто предложить готовые технологические решения, свойственные «новой экономике». Требуется глубоко интегрировать эти решения в существующую на виртуальном рынке сетевую инфраструктуру. Поэтому на смену разработчикам технологических решений в электронной коммерции приходят сетевые посредники, которые реализуют разработанные решения и несут за их эффективность всю полноту ответственности перед партнёрами.

Использование электронной коммерции в качестве базового элемента организации поставок великолепно вписывается в концепцию CSRP (*Customer synchronized resource planning*). Эта концепция предусматривает интеграцию в единую систему управления как движение производственных и материальных ресурсов, так и трансакции, связанные с продвижением товаров и обслуживанием потребителей (См. Рис. 2) [13].

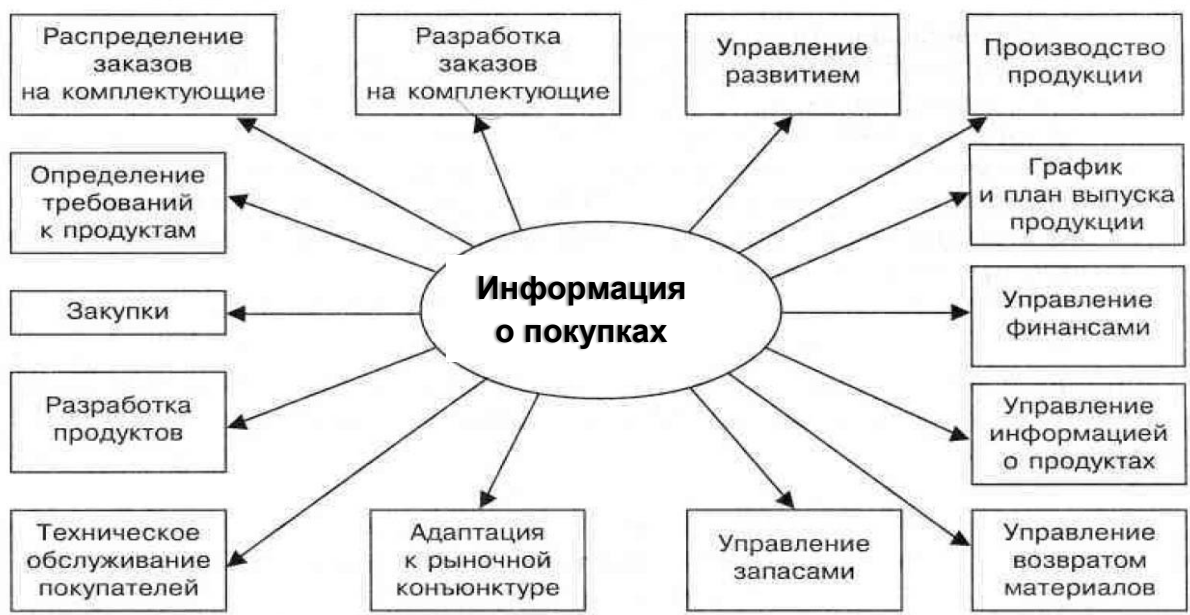

**Рис. 2. Реализация ориентации на интересы покупателя в CSRP-системах.**

Как видно из представленной схемы, информация о покупках является в CSRP-системах отправной точкой принятия всего спектра управленческих решений в цепочке производства и движения товаров. На её основе осуществляется не только управление, но и планирование производственной деятельности предприятия.

Трансформация неформальных институтов товарного рынка создаёт в сетевой экономике то, что Ф. Котлер назвал «сетью несущих ценность отношений» (*Network of value-laden relationships*), когда «*конкурируют уже не столько компании-производители, сколько системы взаимодействия в целом. В этом случае выигрывает та компания, которой удалось построить наиболее эффективную систему*» [4]. Электронная коммерция использует такие системы взаимодействия в качестве механизма обеспечения продаж.

**Управление поставками в электронной коммерции**. Трансформация цепочек поставок в промышленном производстве при переходе к электронной коммерции подразумевает наполнение новым смыслом т.н. «ключевые бизнес-процессы», образующие процесс управления цепочками поставок. Для разъяснения данного тезиса в качестве иллюстрации можно взять стандартную модель логистического управления цепочками поставок, включающую в себя анализ восьми видов бизнес-процессов [15]:

1. *Управление взаимоотношениями с потребителями* в электронной коммерции унифицируется и компьютеризируется. Производитель не взаимодействует с покупателями ни до совершения сделки, ни во время её совершения. Его функции в отношениях с покупателями выполняют электронные посредники. Однако и они ограничены в своих действиях правилами провайдеров услуг электронной коммерции (торговых площадок и платёжных систем).

Торговые провайдеры вырабатывают единые для всех участников электронных сделок правила и обеспечивают безусловное их соблюдение. Выбора у участников торгов нет – любое нарушение правил ведёт к блокированию аккаунта участника. С другой стороны, потребитель заинтересован не в сделке, а в товаре, и чем меньше времени и усилий отнимет у него трансакция, тем привлекательнее будет предстоящая сделка.

У этого есть свои дополнительные плюсы. Торговые провайдеры выступают в качестве институционального посредника, призванного обеспечить баланс интересов участников торгов, сделать так, чтобы ни одна сторона электронной сделки «не тянула одеяло на себя». Их интерес заключается не столько в конкретных сделках (с которых они получают свой процент), сколько в увеличении их числа за счёт создания прозрачных и приемлемых для всех участников торгов правил продаж.

2. ***Управление обслуживанием потребителей*** лишь частично находится в руках товаропроизводителей. Они определяют параметры товаров и ассортиментную политику. От них зависят сроки выполнения заказов и длительность производственного цикла. Причём, управление обслуживанием потребителей в электронной коммерции столь же унифицировано и технологизировано, как и управление взаимоотношениями с клиентами.

Единственное отличие заключается в том, что производитель, благодаря интернет-технологиям, имеет возможность осуществлять обратную связь с потребителями. Речь идёт об использовании горячих телефонных линий, чатов и электронной почты для прямых контактов с клиентами. Для этих же целей используется интернет-регистрация участников электронных торгов, а также размещение информации в общедоступных социальных сетях, интернет-форумах и блогах [11].

В остальном непосредственное обслуживание потребителей осуществляют логистические провайдеры, обеспечивающие непосредственное сопровождение сделок (приём платежей, доставку и т.д.). Их особенность заключается в том, что оказываемые такими посредниками услуги носят унифицированный характер и едины для всех потребителей. Поэтому выбор партнёров в данной области для товаропроизводителей весьма невелик.

3. ***Управление спросом*** в электронной коммерции осуществляется с использованием стандартных инструментов коммуникативных (таргетированная реклама) или торговых (торговые площадки) провайдеров. Речь идёт о достаточно специфических инструментах продвижения, встроенных в функционал торговых площадок и интернет-ресурсов. Например, торговая площадка «Молоток» предлагает продавцам такие универсальные инструменты продвижения как «*Выделение названия лота жирным шрифтом*», «*Подсветка названия лота жёлтым фоном*», «*Добавление в список "Рекомендуемые лоты"*» и ряд других.[4]

Применение традиционных инструментов продвижения в интернет-маркетинге требует больших финансовых затрат, тогда как целевые аудитории в традиционной и электронной коммерции часто не совпадают. Для товаропроизводителей доступны все стандартные инструменты стимулирования продаж. Однако их эффективность невелика из-за необъятности просторов Интернета. Наиболее актуальной здесь является информирующая реклама для массового потребителя и адресная работа с сетевыми посредниками на портале производителя товаров.

4. ***Управление выполнением заказов*** является одной из самых сложных проблем в электронной коммерции. Сложность заключается в том, что речь идёт не столько о перемещении материальных запасов, сколько о виртуальном товародвижении. Это в корне меняет всю систему управления выполнением заказов. Как отмечают М. Кристофер и Х. Пэк: «*При использовании виртуальных запасов товарно-материальные запасы значительно сокращаются, транспортные расходы обычно возрастают, поскольку товары перевозятся в меньших количествах на большие расстояния*» [6]. Собственная транспортная служба товаропроизводителя с такими задачами не справляется. Здесь требуется наличие транспортного провайдера, способного за счёт больших объёмов и географии перевозок обеспечить бесперебойность и своевременность поставок.

Отечественные частные транспортные компании пока не могут решить эту проблему в масштабах всей страны. Поэтому основная нагрузка по обеспечению перевозок ложится на плечи ФГУП «Почта России», работа которого вызывает множество нареканий. Частично проблему может решить РАО «Российские железные дороги», организовавшее электронную информационную площадку для собственников подвижного состава и грузовладельцев.[5] Однако и эта площадка пока не способна обеспечить стабильный грузопоток.

---

[4] Платные услуги и личный счет / Открытая торговая площадка «Molotok.ru». – http://molotok.ru/help_item.php?tid=546&item=3867&zoom=N
[5] Электронная информационная площадка для собственников подвижного состава и грузовладельцев / РАО «РЖД». – http://eip.rzd.ru/Waggon_Exchange/info_list.do

5. ***Управление производственным потоком*** также представляет собой существенную проблему в сетевой экономике, ориентированной на электронную коммерцию. Суть её заключается в том, чтобы трансформировать производственную логистику в «*единый хозяйственный процесс функций транспортировки, управления запасами, разработки новых продуктов, гибкого производства и обслуживания потребителей*». Такая трансформация «*требует перестройки традиционных организационных структур и придания им новых, уникальных конфигураций*» [1].

Сложность состоит в том, что традиционная система производственной логистики ориентирована на эндогенное представление о ёмкости рынка и на текущую пропускную способность сбытовых каналов. Интегрированная с электронной коммерцией логистика, напротив, ориентируется на объём и динамику поступающих заказов, «*сокращая тем самым издержки и стоимость товара для конечного потребителя*» [7]. Поэтому конкурентное преимущество на рынке получает не тот, кто наполнил товарными запасами распределительные сети, а тот, кто обеспечивает своевременное выполнение заказов на основе максимального ускорения производственного цикла.

Пока существуют традиционные каналы товародвижения, преимущества электронной коммерции будут основаны на сопоставимо более низких трансакционных издержках. В условиях экстенсивного роста развитие электронной коммерции происходит за счёт отторжения рынков традиционной оптово-розничной инфраструктуры сбыта. После завершения этого этапа конкуренция переместится в сферу интенсификации и оптимизации производственно-распределительных процессов.

6. ***Управление снабжением*** в сетевой экономике связано с решением сложнейшей задачи «*унификации требований к информации и объединения потоков товарно-материальных запасов для разных уровней поставщиков и потребителей*» [11]. Проблема заключается в формировании сквозной цепочки поставок, включающей все звенья товародвижения – от поставщиков сырья до конечных потребителей товаров. Эталоном может стать модель, в которой поступление заказа запускает всю производственную цепочку вплоть до поставок сырья. На практике всё равно останутся промежуточные товарно-материальные запасы, позволяющие сгладить динамику производства и поставок. Однако конкурентоспособность производителей будет напрямую зависеть от скорости товародвижения и оборачиваемости запасов в цепочке поставок.

Аналог такой системы уже существует на китайской торговой площадке AliExpress под названием «Групповые покупки».[6] Производители заранее определяют дату совершения сделки (с отсрочкой до 30 дней). Затем в течение оставшегося времени торговая площадка аккумулирует предварительные заказы и оплату от покупателей, обмениваясь информацией с производителем, который обеспечивает прямую отгрузку товара покупателям по мелкооптовой цене.

7. ***Управление ассортиментом*** в электронной коммерции подразумевает формирование ассортимента в соответствии с сигналами, получаемыми от контрагентов и потребителей. С одной стороны, это постоянные эксперименты с выпуском на рынок пробных партий продукции, а с другой стороны – формирование механизма интерактивных коммуникаций, вовлекающих торговых партнёров и потребителей в процесс создания и совершенствования товаров.

Электронная коммерция предоставляет производителям уникальные возможности по тестированию рынка через выпуск пробных партий продукции. В отличие от традиционных каналов сбыта размер пробной партии здесь не имеет значения. В виртуальной среде Интернета даже небольшая партия товара сразу становится доступной огромной контактной аудитории. Интерактивные интернет-коммуникации (социальные сети, форумы и блоги) позволяют вовлечь целевую аудиторию в процесс модификации и совершенствования товарного ассортимента. Одновременно это даёт производителям возможность своевре-менно корректировать производственные процессы в соответствии с особенностями спроса и конкуренции на рынке.

---

[6] Групповые покупки / Торговая площадка AliExpress. – http://group.aliexpress.com/ru.htm

8. ***Управление возвратными потоками*** в электронной коммерции делегируется логистическим провайдерам, вовлечённым в процесс товародвижения. Взаимодействуя с потребителями в процессе заключения сделки, они выполняют функции конечного продавца по работе с рекламациями. Задача товаропроизводителей заключается в том, чтобы предложить провайдерам такие условия поставок, которые будут учитывать как возможные претензии покупателей, так и механизм их удовлетворения. Всё это является важным фактором, побуждающим товаропроизводителей обеспечивать должный уровень качества продукции и оказания сервисных услуг (гарантийный ремонт, приём рекламаций, поставка расходных материалов и комплектующих).

Таким образом, независимо от содержания процессов управления поставками их сущность в сетевой экономике сводится к оптимизации трансакционных издержек при организации поставок. Не случайно лауреат Нобелевской премии 1972 года по экономике К. Дж. Эрроу трактует трансакционные издержки как «*затраты на управление экономической системой*» [9]. Основная задача экономических субъектов заключается здесь в такой трансформации цепочек поставок, когда участники электронной коммерции перестанут ощущать недостатки электронных форм организации продаж. От этого зависит сегодня конкурентоспособность российских производителей на глобальном сетевом рынке потребительских товаров. Решение указанной задачи определит ключевое направление не только структурного, но и технологического развития экономики России на ближайшее десятилетие.

*Библиографический список:*
1. *Бауэрсокс Д. Дж., Клосс Д. Дж.* Логистика: интегрированная цепь поставок. М.: Олимп-Бизнес, 2008.
2. *Друкер П.* Эффективное управление. Экономические задачи и оптимальные решения. М.: ФАИР-ПРЕСС, 1998.
3. *Котлер Ф.* 300 ключевых вопросов маркетинга: отвечает Филип Котлер. М.: Олимп-Бизнес, 2006.
4. *Котлер Ф.* Маркетинг менеджмент. СПб.: Питер, 1998.
5. *Коуз Р.* Фирма, рынок и право. М.: Новое издательство, 2007.
6. *Кристофер М., Пэк Х.* Маркетинговая логистика. М.: ИД «Технологии», 2005.
7. *Крымский И. А., Павлов К. В.* Проблемы и перспективы развития электронной экономики в России. Мурманск: Кольский НЦ РАН, 2007.
8. *Сток Дж. Р., Ламберт Д. М.* Стратегическое управление логистикой. М.: Инфра-М, 2005.
9. *Уильямсон О. И.* Экономические институты капитализма: Фирмы, рынки, «отношенческая» контрактация. СПб.: Лениздат, CEV Press, 1996.
10. *Уотерс Д.* Логистика: Управление цепью поставок. М.: ЮНИТИ-ДАНА, 2003.
11. *Хэндфилд Р. Б., Николс Э. Л.* Реорганизация цепей поставок. Создание интегрированных систем формирования ценности. М.: Вильямс, 2003.
12. *Черенков В. И.* Эволюция маркетинговой теории и трансформация доминирующей парадигмы маркетинга. // Вестник СПбУ. Сер. 8. Вып. 2 (16). 2004. С. 3–32.
13. *Юрасов А. В.* Электронная коммерция. М.: Дело, 2003.
14. *Koch R.* The Financial Times guide to strategy: how to create and pursue a winning strategy. 4[th] ed. London: Prentice Hall, 2011.
15. *Lambert D. M., Cooper M. C., Pagh J. D.* Supply Chain Management: Implementation Issues and Research Opportunities. // The International of Journal Logistics Management. Vol. 9 (2). 1998. P. 1–20.